\documentclass{article}
\usepackage{spconf,amsmath,graphicx,hyperref}
\usepackage{cite}
\usepackage{amssymb,amsfonts}
\usepackage{algorithmic}
\usepackage{graphicx}
\usepackage{textcomp}
\usepackage{xcolor}
\usepackage{tabularx}
\usepackage{multirow}
\usepackage{booktabs}
\usepackage{makecell}

\title{Influence of Clean Speech Characteristics on \\ Speech Enhancement Performance}
\name{Mingchi Hou$^{1 ,2}$, Ina Kodrasi$^{1}$\thanks{This work was supported by the Swiss National Science Foundation project 200021\_215187 on ``Pathological Speech Enhancement".}}
\address{$^{1}$Idiap Research Institute, Switzerland \\$^2$École Polytechnique Fédérale de Lausanne, Switzerland}
\def\ninept{\def\baselinestretch{.9}\let\normalsize\small\normalsize}
\begin{document}
\ninept

\maketitle

\begin{abstract}
Speech enhancement (SE) performance is known to depend on noise characteristics and signal-to-noise ratio (SNR), yet intrinsic properties of the clean speech signal itself remain an underexplored factor. In this work, we systematically analyze how clean speech characteristics influence enhancement difficulty across multiple state-of-the-art SE models, languages, and noise conditions. We extract a set of pitch, formant, loudness, and spectral flux features from clean speech and compute correlations with objective SE metrics, including frequency weighted segmental SNR and PESQ. Our results show that formant amplitudes are consistently predictive of SE performance, with higher and more stable formants leading to larger enhancement gains. We further demonstrate that performance varies substantially even within a single speaker’s utterances, highlighting the importance of intra-speaker acoustic variability. These findings provide new insights into SE challenges, suggesting that intrinsic speech characteristics should be considered when designing datasets, evaluation protocols, and enhancement models.
\end{abstract}


\section{Introduction}
Speech enhancement (SE) has seen significant progress in recent years, driven by advances in deep learning and the availability of large-scale corpora. 
Many neural approaches for SE learn direct mappings from noisy to clean speech, commonly through spectral mask prediction~\cite{CNNGRU_mask, NTT2020} or spectral coefficient estimation~\cite{Sun_2017}. While effective, these methods may show limited robustness in unseen noise conditions~\cite{generalization_gap2023}. 
To address these limitations, researchers have explored generative modeling approaches~\cite{richter_speech_2023, lu_diffusion_ddpm_2021, lu_cdiffuse_2022}. 
Generative approaches, such as diffusion-based~\cite{lu_diffusion_ddpm_2021} and Schr{\"o}dinger Bridge-based~\cite{jukic2024sb} models, model speech probabilistically rather than deterministically, improving performance across diverse scenarios.

Although the performance of state-of-the-art SE models has considerably imporoved, the recent URGENT challenge shows that certain acoustic scenarios (such as e.g., wideband or impulse noise) as well as specific speech samples remain particularly challenging to enhance~\cite{Zhang2025Lessons}. These challenges frequently lead to degraded intelligibility and quality, highlighting persistent weaknesses in the design of state-of-the-art models.
As demonstrated in~\cite{Zhang2025Lessons}, a critical issue in this context is the lack of reliable objective indicators of the difficulty of enhancing a given speech sample. 
Traditional analyses have primarily emphasized degradation characteristics such as noise type or signal-to-noise ratio (SNR) as contributors to enhancement difficulty. 
While these aspects are undeniably important, they do not fully characterize the difficulty of enhancing a given speech sample.
For instance, utterances with high SNR values may still contain characteristics that render them hard to process, and even under identical noise conditions and SNRs, different utterances can yield noticeably different results~\cite{Zhang2025Lessons}. 
This mismatch complicates the design of SE training corpora with balanced difficulty distribution, as current measures do not provide sufficient insight into which samples are inherently difficult and why~\cite{Zhang2025Lessons}.

An underexplored perspective is that the difficulty of SE not only depends on the properties of the degradation, but also on intrinsic characteristics of the underlying clean speech signal. Speech is acoustically diverse, with phoneme articulation, temporal dynamics, spectral richness, and voice quality varying significantly across speakers, utterances, and contexts. These intrinsic acoustic features influence how degradations manifest, and consequently, how recoverable the speech is after enhancement.
For example,~\cite{michelsanti_sc_2019} has shown that Lombard speech, a speaking style with altered pitch, loudness, and spectral characteristics, can considerably affect enhancement performance, illustrating the role of intrinsic speech variability.

Speaker-aware SE methods further highlight that intrinsic characteristics of the clean speech signal influence enhancement performance~\cite{TASLP2017,Indiana_WASPAA2021, MS_ICASSP2022, Tencent_ICASSP2022, MS_Interspeech2024, Hou_EUSIPCO_2025}. These methods condition models on speaker embeddings or fine-tune them on speaker-specific data, improving intelligibility and perceptual quality for speakers whose acoustic characteristics are underrepresented in the training data~\cite{Hou_EUSIPCO_2025}. 
These approaches typically assume that variability is primarily among speakers, treating all utterances from a speaker uniformly.
While these results confirm that clean speech characteristics influence SE performance, they do not clarify which specific speech characteristics are most critical, nor how variability within a single speaker's utterances affects enhancement difficulty.

Our work directly investigates which intrinsic acoustic features of clean utterances systematically influence enhancement difficulty by analyzing correlations between acoustic measures and objective SE metrics. 
Specifically, we analyze correlations between SE performance and a set of pitch, formant, loudness, and spectral flux features, extracted from the underlying clean utterances. Our analysis considers multiple SE models, languages, noise types, and SNRs, allowing us to assess whether these intrinsic characteristics consistently relate to enhancement difficulty across different systems and linguistic contexts. By grounding the analysis in clean speech properties rather than degradation conditions alone, we aim to establish a new lens for understanding SE challenges. This perspective provides complementary insights to traditional noise-based measures and can support more principled dataset design and difficulty-aware SE evaluation.

\section{Analysis Framework}

To investigate the relationship between intrinsic speech characteristics and enhancement difficulty, we adopt the following strategy.
We first consider a representative set of state-of-the-art SE models trained following standard practices in the literature (cf. Sections~\ref{sec: se} and~\ref{sec: exp_arch}).
A test set is generated by taking clean speech samples of the same length and creating multiple noisy versions of each sample using a fixed set of noise signals and SNRs (cf. Section~\ref{sec: exp_settings}). 
This ensures that all external factors aside from the intrinsic properties of the speech signals are kept constant across the dataset.
For each clean speech sample in the test set, a set of acoustic features is extracted to characterize it (cf. Section~\ref{sec: acoustic_features}).
SE models are then applied to the noisy signals, and performance improvements are computed using the objective metrics frequency-weighted segmental SNR (fwSSNR)~\cite{fwSSNR} and PESQ~\cite{pesq_2001}.   
For each clean sample, these improvements are averaged across all considered noisy versions of it.
Finally, correlations between the sample-level acoustic features and these average performance improvements are analyzed to identify which intrinsic speech properties systematically influence enhancement difficulty.
To assess whether observed trends generalize across languages, the analysis is separately performed for two languages, with models trained and tested on English and models trained and tested on Spanish.
In the following, we briefly review the considered SE systems and provide a more detailed description of the acoustic features used.

\subsection{Speech enhancement models}
\label{sec: se}

We analyze a representative set of state-of-the-art SE models that span different modeling paradigms, including mask-based prediction, complex regression, diffusion, and optimal transport.
In the following, we briefly summarize the systems considered in our analysis.  
Details on the implementation and training of these models are presented in Section~\ref{sec: exp_settings}.

\emph{Magnitude spectrogram masking (MM).} \enspace 
Mask-based models enhance speech by selectively suppressing noise-dominated time–frequency components in the noisy spectrogram~\cite{vincent_spectral_2018}.  
Our MM system uses a deep neural network (DNN) to predict an ideal ratio mask, constrained to $[0,1]$ via a sigmoid activation.  
The model is optimized with the scale-invariant signal-to-distortion ratio~\cite{roux_sdr_2019}.  

\emph{Complex spectrogram regression (CR).} \enspace 
While MM models leave the noisy phase unaltered, regression-based methods directly estimate both magnitude and phase.  
Our CR system uses a DNN that predicts the real and imaginary parts of the clean short-time Fourier transform (STFT) coefficients~\cite{song2019generative}, trained with a time-domain mean square error (MSE) loss against the clean reference.  

\emph{Score-based diffusion (SGMSE+).} \enspace 
Diffusion models enhance speech by reversing a gradual noise corruption process~\cite{song_2021_score}.  
Our SGMSE+ system trains a DNN to estimate the score function that guides this reverse process, augmented with an affine drift term and a predictor–corrector sampler for improved stability and reconstruction quality~\cite{richter2023sgmse}.  

\emph{Schr\"{o}dinger Bridge (SB).} \enspace 
The SB formulation casts SE as an optimal transport problem between noisy and clean speech distributions~\cite{chen2023sb, jukic2024sb}.  
Instead of diffusing clean signals toward noise, the SB approach learns an interpolation that directly couples noisy inputs with their clean counterparts.  
Our SB system employs an SDE-based sampler to generate enhanced waveforms~\cite{jukic2024sb}.

\subsection{Acoustic features}
\label{sec: acoustic_features}
To investigate how intrinsic speech characteristics may influence enhancement difficulty, we extract a set of acoustic features using the openSMILE toolkit~\cite{os}.
In the following, we describe the considered acoustic features and the rationale behind their selection.

\emph{Pitch.} \enspace 
Pitch characterizes the harmonic structure of an utterance and its prosodic patterns, which can influence enhancement performance.  
Utterances with low pitch may be challenging to enhance because their harmonics are closely spaced and more easily masked by noise.  
Similarly, low pitch variability (i.e., monotone speech) reduces temporal and spectral cues, potentially making it harder for models to distinguish speech from background noise.
To quantify pitch characteristics for an utterance, we consider the mean and standard deviation of the fundamental frequency $f_0$ extracted using robust tracking and smoothing~\cite{eyben2016geneva}.  

\emph{Formants.} \enspace
Formants describe the resonant frequencies of the vocal tract and shape the spectral envelope of speech, providing important spectral cues.  
Low formant amplitudes can reduce the local SNR, making speech harmonics more easily masked by noise.  
High variability in formant amplitudes indicates rapidly fluctuating spectral peaks, which may be difficult for SE models to track and reconstruct accurately.  
Conversely, low variability indicates flat formants, potentially providing fewer dynamic cues to distinguish speech from noise.
To quantify formant characteristics, we consider the mean and standard deviation of the amplitudes of the first three formants ($F_1$, $F_2$, $F_3$) extracted using linear predictive coding and peak-picking in the smoothed spectrum~\cite{eyben2016geneva}.  

\emph{Loudness.} \enspace 
Loudness characterizes the perceived intensity of speech.
Utterances with low loudness levels may be challenging for SE models, because low-energy segments can be easily masked by noise.  
Furthermore, high loudness variability within an utterance indicates rapid intensity fluctuations, which may be difficult for models to track.
Conversely, low loudness variability may provide fewer dynamic cues, potentially reducing the ability of SE models to distinguish speech from noise.
To quantify loudness characteristics, we consider the utterance-level mean and standard deviation of the loudness level computed using a psychoacoustic model with short-term smoothing~\cite{eyben2016geneva}.


\emph{Spectral flux.} \enspace 
Spectral flux measures the frame-to-frame change of the short-term spectrum, reflecting how the spectral envelope evolves over time.  
Unlike formant amplitudes, which describe static resonances at a given time, spectral flux captures the dynamics of those resonances and other spectral components.  
Utterances with high spectral flux, for instance due to rapid consonant–vowel transitions or expressive articulation, present frequent and abrupt changes that may challenge SE models.  
Conversely, low spectral flux indicates relatively stable spectra, which may provide fewer dynamic cues and challenge SE models when the noise spectrum overlaps strongly with speech.  
To capture these dynamics, we compute both the mean and standard deviation of spectral flux across each utterance as in~\cite{eyben2016geneva}.  


\section{Experimental settings}
\label{sec: exp_settings}
\subsection{Datasets}

{\emph{Clean speech datasets.}} \enspace
To assess the generalizability of our findings across languages, we use two clean speech datasets, i.e., the English Wall Street Journal (WSJ0) dataset~\cite{garofolo_john_s_csr-i_2007_wsj0} and the Crowdsourced Latin American Spanish (CROWD) dataset~\cite{guevara-rukoz_crowdsourcing_2020}. WSJ0 is a widely adopted benchmark in SE research~\cite{richter2023sgmse, jukic2024sb}, comprising $120$ speakers and $28.6$ hours of recordings. For training and evaluation, we follow the official partitioning into $101$ training, $10$ validation, and $8$ test speakers. Although CROWD is larger than WSJ0, we use a subset of it to ensure a fair comparison. This subset is constructed so that the training, validation, and test sets contain the same number of speakers as in WSJ0, totaling $26.7$ hours of audio. All signals are downsampled to $16$ kHz.
To ensure that test performance is not influenced by varying signal lengths and depends only on the intrinsic properties of the clean signals, the available test utterances are segmented into $2$ second chunks, and $200$ chunks are randomly selected per speaker. This procedure results in $1,600$ clean test samples per dataset (i.e., $200$ chunks from each of the $8$ test speakers).

{\emph{Noisy mixtures.}} \enspace To generate noisy mixtures, we use four noise types (bus, cafe, pedestrian area, and street) from the CHiME3 dataset~\cite{barker_chime3_2015}. All noise signals are first downsampled to $16$ kHz. Following common practice in the literature~\cite{lemercier_storm_2023,jukic2024sb}, the noisy training and validation sets are created by randomly selecting a noise file and adding it to a clean signal with an SNR randomly chosen between $-6$ dB and $14$ dB.
For the noisy test set, to ensure that performance depends only on the intrinsic properties of the clean signals and is not influenced by varying noise types or SNRs, multiple noisy versions of each clean test sample are generated using the fixed set of three SNRs ($-5$ dB, $5$ dB, and $15$ dB) and the same $2$ second chunk from each of the four noise types. This procedure results in $19200$ noisy test samples per dataset (i.e., $12$ noisy versions for each of the $1600$ clean test samples).




\subsection{Model architectures and training}
\label{sec: exp_arch}
As in~\cite{lemercier_storm_2023}, signals are transformed to the STFT domain using a window size of $510$ samples and a hop size of $128$ samples. To compress the dynamic range, the STFT coefficients are further processed with $\alpha = 0.5$ and $\beta = 0.33$ as in~\cite{jukic2024sb}.

\mbox{MM} is implemented using a $5$-layer BiLSTM~\cite{NTT2020}. \mbox{CR} follows the NCSN+ U-Net architecture, modified for complex inputs as in~\cite{lemercier_storm_2023}. \mbox{SGMSE+} uses denoising score matching with hyperparameters $\sigma_{\min}=0.05$, $\sigma_{\max}=0.5$, $\gamma=1.5$, and a $30$-step predictor–corrector sampler~\cite{richter2023sgmse}. \mbox{SB} uses a variance-exploding schedule with hyperparameters $\sigma_{\min}=0.7$, $\sigma_{\max}=1.82$, and $50$-step SDE sampling. Both \mbox{SGMSE+} and \mbox{SB} use NCSN+ backbones with noise-scheduling layers and exponential moving average with a weight decay of $0.999$~\cite{jukic2024sb}. The number of trainable parameters for \mbox{MM}, \mbox{CR}, \mbox{SGMSE+}, and \mbox{SB} is $7.6$ million, $22.1$ million, $25.2$ million, and $25.2$ million, respectively.

Training is performed using the Adam optimizer with a batch size of $8$, a learning rate of $10^{-4}$, and a maximum of $1000$ epochs. Training stops if the validation loss does not decrease for $20$ consecutive epochs. The \mbox{CR}, \mbox{SGMSE+}, and \mbox{SB} models are trained on an NVIDIA H100 GPU, whereas \mbox{MM} is trained on an RTX 3090 GPU.



\begin{table*}[t!]
\footnotesize
    \centering
    \caption{Pearson correlation coefficients between $\Delta$fwSSNR of the considered SE models and acoustic features. Models are separately trained and tested on English (En.) and Spanish (Sp.) datasets. $\mu$ and $\sigma$ denote the mean and standard deviation of the considered acoustic features. Values in bold indicate the highest correlation for each model and dataset. Statistically insignificant correlations are grayed.}
    \label{tbl: corrs_fwssnr}
\begin{tabularx}{\textwidth}{|ll||*{6}{>{\centering\arraybackslash}X|>{\centering\arraybackslash}X|}}
\toprule
\multicolumn{1}{|l}{\multirow{2}{*}{Model}} & 
\multicolumn{1}{l||}{\multirow{2}{*}{Lang.}}
    & \multicolumn{2}{c|}{$f_0$} 
    & \multicolumn{2}{c|}{$F_1$} 
    & \multicolumn{2}{c|}{$F_2$} 
    & \multicolumn{2}{c|}{$F_3$} 
    & \multicolumn{2}{c|}{Loudness} 
    & \multicolumn{2}{c|}{Flux} \\
\cmidrule(lr){3-4} \cmidrule(lr){5-6} \cmidrule(lr){7-8} \cmidrule(lr){9-10} 
\cmidrule(lr){11-12} \cmidrule(lr){13-14} 
\multicolumn{2}{|c||}{} 
    & $\mu$ & $\sigma$ & $\mu$ & $\sigma$ & $\mu$ & $\sigma$ & $\mu$ & $\sigma$ & $\mu$ & $\sigma$ & $\mu$ & $\sigma$ \\
\midrule
\multirow{2}{*}{MM} & En. & $0.50$ & \textcolor{gray}{$-0.05$} & $0.65$ & $-0.61$ & $0.68$ & $-0.64$ & $\bf 0.69$ & $-0.66$ & $0.37$ & $-0.58$ & $0.20$ & $-0.52$ \\
                     & Sp. & $0.28$ & $0.16$ & $0.62$ & $-0.60$ & $0.64$ & $-0.63$ & $\bf 0.65$ & $-0.61$ & $0.43$ & $-0.46$ &  $0.34$ & $-0.48$ \\
\midrule
\multirow{2}{*}{CR} & En. & $0.41$ & \textcolor{gray}{$-0.03$} & $0.75$ & $-0.73$ & $0.77$ & $-0.74$ & $\bf 0.78$ & $-0.77$ & $0.49$ & $-0.65$ & $0.30$ & $-0.59$ \\
                     & Sp. & $0.21$ & $0.06$ & $0.70$ & $-0.67$ & $\bf 0.72$ & $-0.71$ & $\bf 0.72$ & $-0.70$ & $0.60$ & $-0.52$ &  $0.48$ & $-0.48$ \\
\midrule
\multirow{2}{*}{SGMSE+} & En. & $0.37$ & \textcolor{gray}{$-0.03$} & $0.68$ & $-0.67$ & $0.70$ & $-0.67$ & $\bf 0.72$ & $-0.71$ & $0.46$ & $-0.64$ & $0.26$ & $-0.59$ \\
                         & Sp. & $0.17$ & \textcolor{gray}{$0.05$} & $0.51$ & $\bf -0.54$ & $0.51$ & $-0.50$ & $0.51$ & $-0.46$ & $0.50$ & $-0.38$  & $0.50$ & $-0.28$ \\
\midrule
\multirow{2}{*}{SB} & En. & $0.38$ & \textcolor{gray}{$-0.01$} & $0.70$ & $-0.68$ & $0.72$ & $-0.69$ & $\bf 0.74$ & $-0.72$ & $0.48$ & $-0.66$ & $0.28$ & $-0.60$ \\
                     & Sp. & $\bf 0.45$ & \textcolor{gray}{$0.08$} & $0.30$ & $-0.33$ & $0.34$ & $-0.37$ & $0.35$ & $-0.37$ & $0.30$ & $-0.10$ &  $0.27$ & $-0.16$ \\
\bottomrule
\end{tabularx}
\end{table*}

\begin{table*}[t!]
\footnotesize
    \centering
    \caption{Pearson correlation coefficients between $\Delta$PESQ of the considered SE models and acoustic features. Models are separately trained and tested on English (En.) and Spanish (Sp.) datasets. $\mu$ and $\sigma$ denote the mean and standard deviation of the considered acoustic features. Values in bold indicate the highest correlation for each model and dataset. Statistically insignificant correlations are grayed.}
    \label{tbl: corrs_pesq}
\begin{tabularx}{\textwidth}{|ll||*{6}{>{\centering\arraybackslash}X|>{\centering\arraybackslash}X|}}
\toprule
\multicolumn{1}{|l}{\multirow{2}{*}{Model}} & 
\multicolumn{1}{l||}{\multirow{2}{*}{Lang.}}
    & \multicolumn{2}{c|}{$f_0$} 
    & \multicolumn{2}{c|}{$F_1$} 
    & \multicolumn{2}{c|}{$F_2$} 
    & \multicolumn{2}{c|}{$F_3$} 
    & \multicolumn{2}{c|}{Loudness} 
    & \multicolumn{2}{c|}{Flux} \\
\cmidrule(lr){3-4} \cmidrule(lr){5-6} \cmidrule(lr){7-8} \cmidrule(lr){9-10} 
\cmidrule(lr){11-12} \cmidrule(lr){13-14} 
\multicolumn{2}{|c||}{} 
    & $\mu$ & $\sigma$ & $\mu$ & $\sigma$ & $\mu$ & $\sigma$ & $\mu$ & $\sigma$ & $\mu$ & $\sigma$ & $\mu$ & $\sigma$ \\
\midrule
\multirow{2}{*}{MM} & En. & $0.39$ & $0.09$ & $0.36$ & $-0.34$ & $0.40$ & $-0.39$ & $\bf 0.41$ & $\bf-0.41$ & $0.31$ & $-0.22$ & $0.21$ & $-0.22$ \\
                     & Sp. & $\bf 0.51$ & \textcolor{gray}{$0.04$} & $-0.07$ & \textcolor{gray}{$0.05$} & \textcolor{gray}{$-0.03$} & \textcolor{gray}{$-0.02$} & \textcolor{gray}{$-0.03$} & \textcolor{gray}{$-0.06$} & \textcolor{gray}{$-0.06$} & $0.32$ &  $-0.11$ & \textcolor{gray}{$0.08$} \\
\midrule
\multirow{2}{*}{CR} & En. & $0.46$ & \textcolor{gray}{$0.04$} & $0.47$ & $-0.45$ & $0.52$ & $-0.51$ & $\bf 0.53$ & $-0.52$ & $0.33$ & $-0.29$ & $0.20$ & $-0.27$ \\
                     & Sp. & $\bf 0.54$ & \textcolor{gray}{$-0.06$} & $0.16$ & $-0.12$ & $0.21$ & $-0.26$ & $0.22$ & $-0.33$ & $0.12$ & \textcolor{gray}{$0.05$} & \textcolor{gray}{$-0.05$} & $-0.18$ \\
\midrule
\multirow{2}{*}{SGMSE+} & En. & $0.41$ & \textcolor{gray}{$0.01$} & $0.41$ & $-0.40$ & $0.45$ & $-0.44$ & $\bf 0.46$ & $\bf -0.46$ & $0.30$ & $-0.30$ & $0.16$ & $-0.27$ \\
                         & Sp. & $0.24$ & \textcolor{gray}{$-0.03$} & $0.24$ & $-0.21$ & $0.26$ & $-0.29$ & $0.26$ & $\bf -0.32$ & $0.29$ & \textcolor{gray}{$-0.06$} & $0.21$ & $-0.22$ \\
\midrule
\multirow{2}{*}{SB} & En. & $0.47$ & \textcolor{gray}{$0.001$} & $0.42$ & $-0.41$ & $\bf 0.48$ & $-0.46$ & $ \bf 0.48$ & $\bf -0.48$ & $0.32$ & $-0.28$  & $0.20$ & $-0.25$ \\
                     & Sp. & $\bf 0.42$ & \textcolor{gray}{$0.02$} & \textcolor{gray}{$-0.04$} & \textcolor{gray}{$0.03$} & \textcolor{gray}{$-0.01$} & \textcolor{gray}{$-0.05$} & \textcolor{gray}{$-0.003$} & $-0.09$ & \textcolor{gray}{$-0.03$} & $0.29$ & $-0.11$ & \textcolor{gray}{$0.06$} \\
\bottomrule
\end{tabularx}
\end{table*}
\begin{table*}[t!]
\footnotesize
\centering
\caption{Performance (mean $\pm$ standard deviation) of the considered SE models on English and Spanish utterances in the highest ($Q_4$) and lowest ($Q_1$) quartiles of mean $F_3$ values.}
\label{tbl:perf_feat}
\setlength{\tabcolsep}{3pt}
\begin{tabularx}{\textwidth}{|c|c||*{8}{>{\centering\arraybackslash}X|}@{}}
\toprule
\multirow{2}{*}{Measure} & \multirow{2}{*}{Quartile} 
     & \multicolumn{2}{c|}{MM} 
     & \multicolumn{2}{c|}{CR} 
     & \multicolumn{2}{c|}{SGMSE+} 
     & \multicolumn{2}{c|}{SB} \\
 \cmidrule(lr){3-4} \cmidrule(lr){5-6} \cmidrule(lr){7-8} \cmidrule(lr){9-10} 
  & & English & Spanish & English & Spanish & English & Spanish & English & Spanish \\
 \midrule
 \multirow{2}{*}{$\Delta$fwSSNR [dB]} 
     & $Q_4$ & $4.35 \pm 1.13$ & $4.19 \pm 1.18$ & $7.01 \pm 1.11$ & $6.03 \pm 1.32$ & $6.10 \pm 0.96$ & $4.76 \pm 0.93$ & $8.06 \pm 1.12$ & $5.75 \pm 1.14$ \\
     & $Q_1$ & $1.86 \pm 0.96$ & $2.14 \pm 1.15$ & $3.93 \pm 1.03$ & $3.52 \pm 0.98$ & $3.84 \pm 0.81$ & $3.60 \pm 0.81$ & $5.37 \pm 0.97$ & $4.92 \pm 1.30$ \\
 \midrule
 \multirow{2}{*}{$\Delta$PESQ} 
     & $Q_4$ & $1.10 \pm 0.13$ & $0.90 \pm 0.15$ & $1.46 \pm 0.15$ & $1.25 \pm 0.19$ & $1.00 \pm 0.13$ & $0.63 \pm 0.13$ & $1.59 \pm 0.18$ & $1.02 \pm 0.19$ \\
     & $Q_1$ & $0.91 \pm 0.24$ & $0.98 \pm 0.40$ & $1.14 \pm 0.29$ & $1.14 \pm 0.39$ & $0.78 \pm 0.21$ & $0.52 \pm 0.27$ & $1.29 \pm 0.27$ & $1.08 \pm 0.42$ \\
 \bottomrule
 \end{tabularx}
 \end{table*}
\subsection{Evaluation and analysis}

SE performance is evaluated using two objective metrics, i.e., fwSSNR~\cite{fwSSNR} and PESQ~\cite{torcoli_pesqc2_2025}. For both metrics, the clean speech signal is used as the reference signal. To facilitate comparison, we consider the difference between the metrics of the enhanced signal and the noisy mixtures, denoted as $\Delta$fwSSNR and $\Delta$PESQ.

For each clean test sample, the corresponding acoustic features described in Section~\ref{sec: acoustic_features} are computed. For each clean sample there are $12$ noisy versions, and the improvements $\Delta$fwSSNR and $\Delta$PESQ are averaged across these versions. To quantify the relationship between acoustic features and SE performance, Pearson correlation coefficients are calculated between each acoustic feature and the averaged metric values. 
The statistical significance of the correlation values is assessed using a two-sided t-test with a p-value threshold of $0.001$~\cite{cohen_statistical_2009}.

\section{Experimental Results}
\subsection{Correlation between acoustic features and SE performance}
\label{sec: exp_corrs}
Table~\ref{tbl: corrs_fwssnr} presents the correlation coefficients between the considered acoustic features and $\Delta$fwSSNR for all models and both datasets. Overall, formant-related features (i.e., $F_1$, $F_2$, and $F_3$) show the strongest correlations with $\Delta$fwSSNR across models and languages.
Mean formant values are positively correlated with performance improvements, while their standard deviations are negatively correlated. This indicates that speech signals with strong and stable formant amplitudes are considerably easier for SE models to enhance. In particular, the mean of $F_3$ typically exhibits the strongest correlation, though $F_1$ and $F_2$ also show similarly strong correlations.
Other acoustic features also reveal meaningful patterns, particularly in English. Signals with stable loudness and spectral dynamics tend to enhance more effectively, as reflected in strong negative correlations of $\Delta$fwSSNR with their standard deviations. While mean $f_0$ shows moderate effects for some models, pitch variability has little impact, suggesting SE performance relies more on overall pitch than expressive cues. Finally, it can be observed that correlations are generally stronger in English than Spanish, though the role of formant amplitudes remains consistent, and these trends persist across noise types and SNR levels.


Table~\ref{tbl: corrs_pesq} shows the correlations between acoustic features and $\Delta$PESQ. Overall, correlations are weaker for $\Delta$PESQ than for $\Delta$fwSSNR. However, for the English dataset, the same general trends are observed, with F$_2$ and F$_3$ showing the strongest correlations across models. In Spanish however, the mean of $f_0$ tends to be the most relevant predictor, while correlations with formant amplitudes are often weak. This indicates a language-dependent effect, where PESQ improvements are more reliably predicted by acoustic features in English than in Spanish.
We suspect this occurs due to the suboptimality in using PESQ for predicting perceptual quality of Spanish samples, given that it was not optimized for Spanish~\cite{itu_p8621, konane_impact_lang_2021}.

\begin{figure}[b!]
\begin{minipage}[b]{.48\linewidth}
  \centering
  \centerline{\includegraphics[width=4.7cm]{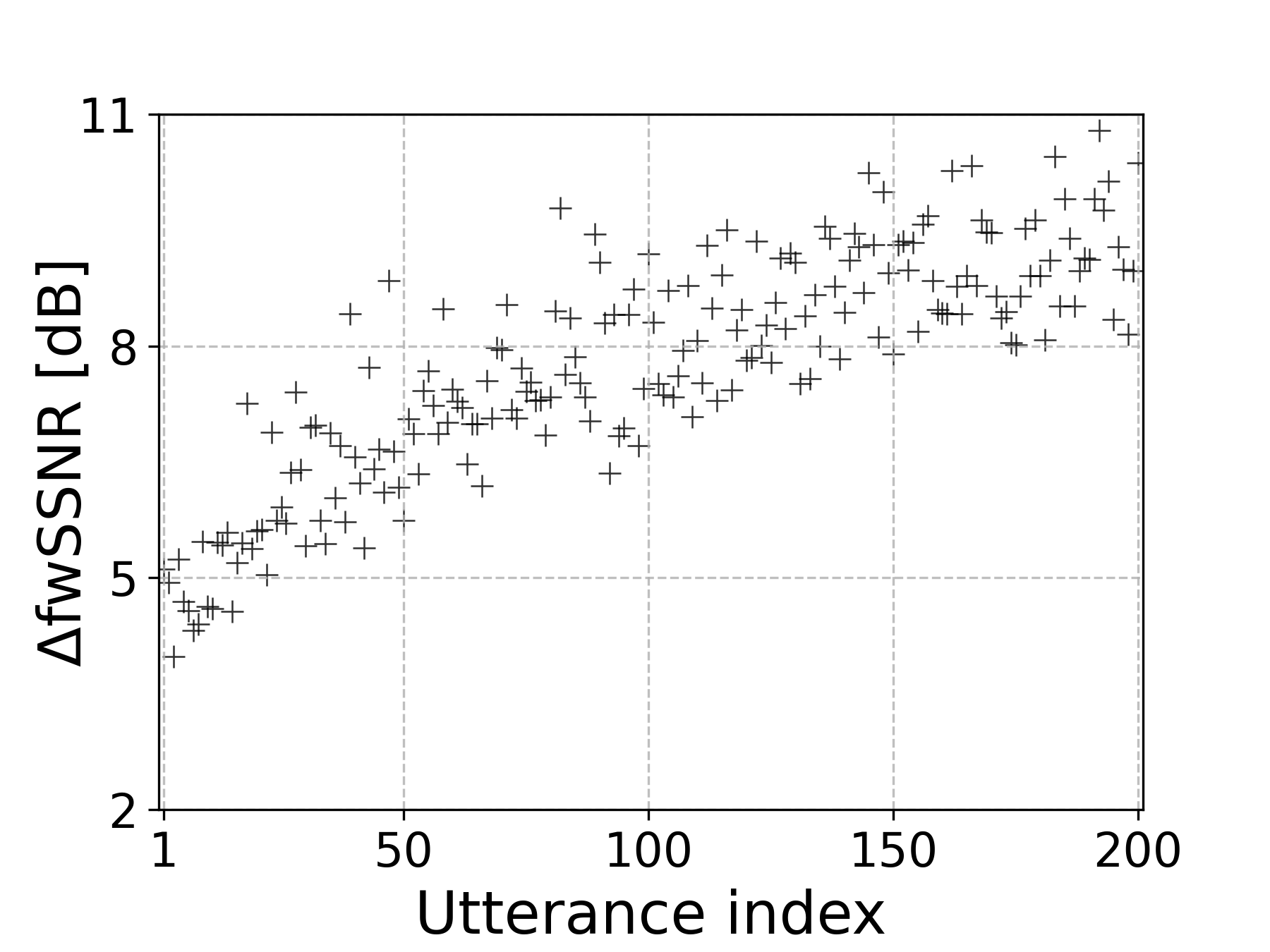}}
  \centerline{(a) English Speaker}
\end{minipage}
\hfill
\begin{minipage}[b]{0.48\linewidth}
  \centering
  \centerline{\includegraphics[width=4.7cm]{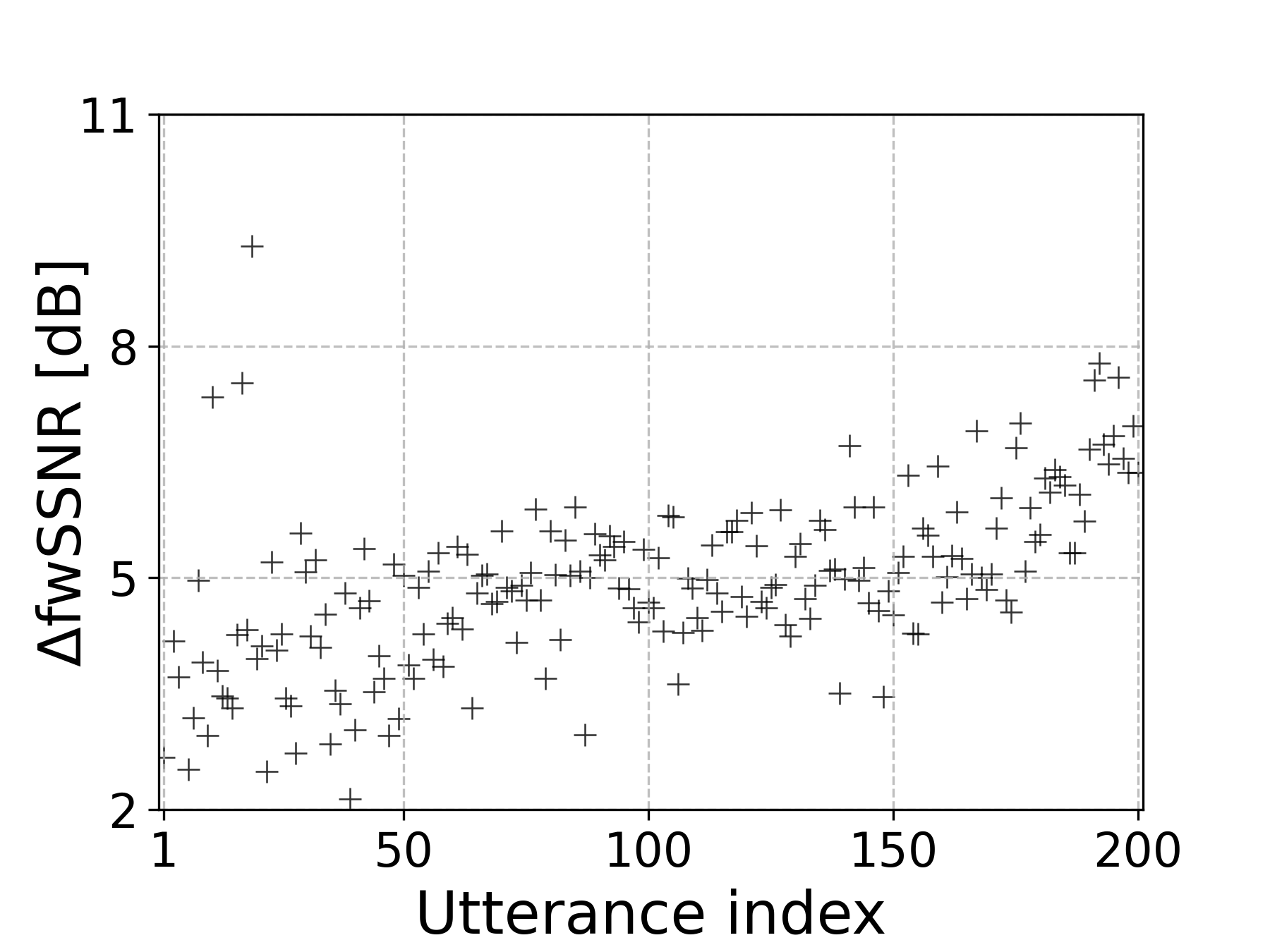}}
  \centerline{(b) Spanish Speaker}
\end{minipage}
\caption{$\Delta$fwSSNR of the SB model across all utterances from (a) an exemplary English speaker and (b) an exemplary Spanish speaker. Utterances are sorted by increasing mean $F_3$ values.}
\label{fig:2spkr_scatter_en}
\end{figure}
\subsection{SE performance across acoustic feature ranges}
Since correlations alone do not reveal the magnitude of performance differences, in this section we provide a more detailed analysis of how performance varies for utterances with different acoustic characteristics. Because $F_3$ showed the strongest correlations with SE performance, we selected it for this analysis. We divided the test samples into two groups based on the first and fourth quartiles of their mean $F_3$ values. The first group $Q_1$ includes samples in the lowest 25\% of mean $F_3$ values, and the second group $Q_4$ includes samples in the highest 25\%. For each group, we computed the average performance improvements in terms of $\Delta$fwSSNR and $\Delta$PESQ. The results for all considered models and both datasets are presented in Table~\ref{tbl:perf_feat}.
It can be observed that utterances with higher mean $F_3$ are considerably easier to enhance compared to those with lower mean $F_3$ values. In terms of $\Delta$fwSSNR, all models exhibit performance gains of $2$ dB-$3$ dB for samples in $Q_4$ relative to $Q_1$ across both languages. 
In terms of $\Delta$PESQ, utterances in $Q_4$ achieve $0.2$-$0.3$ higher scores than those in $Q_1$ for English, highlighting a clear perceptual advantage. 
Consistent with the results presented in Section~\ref{sec: exp_corrs}, the Spanish dataset shows smaller differences in perceptual quality, with improvements between quartiles typically around $0.1$ points. We attribute this to the limited reliability of PESQ for Spanish~\cite{konane_impact_lang_2021}.

\subsection{Performance variability of utterances from the same speaker}
Our previous results highlighted that intrinsic acoustic features, particularly formant amplitudes, systematically influence enhancement difficulty. While speaker-aware SE research often concentrates on inter-speaker variability, these findings suggest that performance may also vary substantially across utterances from the same speaker.
To examine the role of within-speaker variability, we analyze enhancement performance across different utterances from the same speaker using the SB model as an illustrative example. Fig.~\ref{fig:2spkr_scatter_en} presents scatter plots of the obtained $\Delta$fwSSNR values for all utterances from one exemplary English speaker and one exemplary Spanish speaker. It should be noted that utterances are sorted by their mean $F_3$ values.
The results clearly show that enhancement performance can vary substantially across utterances, even when produced by the same speaker. For both speakers, $\Delta$fwSSNR values span a wide range, demonstrating that utterance-level acoustic variability is a critical factor in SE performance. Since we sorted utterances by their mean $F_3$ values, part of this variability can be explained by differences in mean $F_3$ across utterances. This observation complements the speaker-aware SE literature by highlighting that intra-speaker variability can be as influential as inter-speaker differences.

\section{Conclusion and Outlook}
This paper showed that intrinsic acoustic features of clean speech systematically influence SE performance. By analyzing correlations with features such as pitch, formants, loudness, and spectral flux, we identified formant amplitudes as the most consistent predictor of enhancement gains. Our analysis also revealed that utterances with stable loudness and spectral flux are easier to enhance, while high within-utterance variability can reduce performance. These insights highlight the importance of considering intrinsic speech characteristics, alongside noise conditions, when evaluating, training, or benchmarking SE systems. 

\footnotesize
\bibliographystyle{IEEEtran}
\bibliography{refs}

\begin{thebibliography}{10}
\providecommand{\url}[1]{#1}
\csname url@samestyle\endcsname
\providecommand{\newblock}{\relax}
\providecommand{\bibinfo}[2]{#2}
\providecommand{\BIBentrySTDinterwordspacing}{\spaceskip=0pt\relax}
\providecommand{\BIBentryALTinterwordstretchfactor}{4}
\providecommand{\BIBentryALTinterwordspacing}{\spaceskip=\fontdimen2\font plus
\BIBentryALTinterwordstretchfactor\fontdimen3\font minus \fontdimen4\font\relax}
\providecommand{\BIBforeignlanguage}[2]{{%
\expandafter\ifx\csname l@#1\endcsname\relax
\typeout{** WARNING: IEEEtran.bst: No hyphenation pattern has been}%
\typeout{** loaded for the language `#1'. Using the pattern for}%
\typeout{** the default language instead.}%
\else
\language=\csname l@#1\endcsname
\fi
#2}}
\providecommand{\BIBdecl}{\relax}
\BIBdecl

\bibitem{CNNGRU_mask}
M.~Hasannezhad, Z.~Ouyang, W.-P. Zhu, and B.~Champagne, ``An integrated cnn-gru framework for complex ratio mask estimation in speech enhancement,'' in \emph{Asia-Pacific Signal and Information Processing Association Annual Summit and Conference (APSIPA ASC)}, 2020, pp. 764--768.

\bibitem{NTT2020}
K.~Kinoshita, T.~Ochiai, M.~Delcroix, and T.~Nakatani, ``Improving noise robust automatic speech recognition with single-channel time-domain enhancement network,'' in \emph{Proc. IEEE Int. Conf. on Acoustics, Speech and Signal Processing (ICASSP)}, 2020, pp. 7009--7013.

\bibitem{Sun_2017}
L.~Sun, J.~Du, L.-R. Dai, and C.-H. Lee, ``Multiple-target deep learning for lstm-rnn based speech enhancement,'' in \emph{2017 Hands-free Speech Communications and Microphone Arrays (HSCMA)}, 2017, pp. 136--140.

\bibitem{generalization_gap2023}
P.~Gonzalez, T.~S. Alstr\o{}m, and T.~May, ``Assessing the generalization gap of learning-based speech enhancement systems in noisy and reverberant environments,'' \emph{IEEE/ACM Trans. Audio, Speech and Lang. Proc.}, vol.~31, p. 3390–3403, Sep. 2023.

\bibitem{richter_speech_2023}
J.~Richter, S.~Welker, J.-M. Lemercier, B.~Lay, and T.~Gerkmann, ``Speech {Enhancement} and {Dereverberation} {With} {Diffusion}-{Based} {Generative} {Models},'' \emph{IEEE/ACM Transactions on Audio, Speech, and Language Processing}, vol.~31, pp. 2351--2364, 2023.

\bibitem{lu_diffusion_ddpm_2021}
Y.-J. Lu, Y.~Tsao, and S.~Watanabe, ``A {Study} on {Speech} {Enhancement} {Based} on {Diffusion} {Probabilistic} {Model},'' \emph{Proc. (APSIPA ASC)}, 2021.

\bibitem{lu_cdiffuse_2022}
Y.-J. Lu, Z.-Q. Wang, S.~Watanabe, A.~Richard, C.~Yu, and Y.~Tsao, ``Conditional {Diffusion} {Probabilistic} {Model} for {Speech} {Enhancement},'' in \emph{Proc. IEEE Int. Conf. on Acoustics, Speech and Signal Processing (ICASSP)}, May 2022, pp. 7402--7406.

\bibitem{jukic2024sb}
A.~Juki{\'c}, R.~Korostik, J.~Balam, and B.~Ginsburg, ``Schrödinger bridge for generative speech enhancement,'' in \emph{Proc. Interspeech}, 2024.

\bibitem{Zhang2025Lessons}
W.~Zhang, K.~Saijo, S.~Cornell, R.~Scheibler, C.~Li, Z.~Ni, A.~Kumar, M.~Sach, W.~Wang, Y.~Fu, S.~Watanabe, T.~Fingscheidt, and Y.~Qian, ``Lessons learned from the urgent 2024 speech enhancement challenge,'' in \emph{Proc. Interspeech}, Rotterdamn, The Netherlands, Aug. 2025, pp. 853--857.

\bibitem{michelsanti_sc_2019}
D.~Michelsanti, Z.-H. Tan, S.~Sigurdsson, and J.~Jensen, ``Deep-learning-based audio-visual speech enhancement in presence of lombard effect,'' \emph{Speech Communication}, vol. 115, pp. 38--50, Dec. 2019.

\bibitem{TASLP2017}
M.~Kolbæk, Z.-H. Tan, and J.~Jensen, ``Speech intelligibility potential of general and specialized deep neural network based speech enhancement systems,'' \emph{IEEE/ACM Transactions on Audio, Speech, and Language Processing}, vol.~25, no.~1, pp. 153--167, 2017.

\bibitem{Indiana_WASPAA2021}
S.~Kim and M.~Kim, ``Test-time adaptation toward personalized speech enhancement: Zero-shot learning with knowledge distillation,'' in \emph{IEEE Workshop on Applications of Signal Processing to Audio and Acoustics (WASPAA)}, 2021, pp. 176--180.

\bibitem{MS_ICASSP2022}
S.~E. Eskimez, T.~Yoshioka, H.~Wang, X.~Wang, Z.~Chen, and X.~Huang, ``Personalized speech enhancement: new models and comprehensive evaluation,'' in \emph{Proc. IEEE Int. Conf. on Acoustics, Speech and Signal Processing (ICASSP)}, 2022, pp. 356--360.

\bibitem{Tencent_ICASSP2022}
Y.~Ju, W.~Rao, X.~Yan, Y.~Fu, S.~Lv, L.~Cheng, Y.~Wang, L.~Xie, and S.~Shang, ``Tea-pse: Tencent-ethereal-audio-lab personalized speech enhancement system for icassp 2022 dns challenge,'' in \emph{Proc. IEEE Int. Conf. on Acoustics, Speech and Signal Processing (ICASSP)}, 2022, pp. 9291--9295.

\bibitem{MS_Interspeech2024}
T.~Pärnamaa and A.~Saabas, ``Personalized {Speech} {Enhancement} {Without} a {Separate} {Speaker} {Embedding} {Model},'' in \emph{Proc. Interspeech}.\hskip 1em plus 0.5em minus 0.4em\relax ISCA, Sep. 2024, pp. 4863--4867.

\bibitem{Hou_EUSIPCO_2025}
M.~Hou and I.~Kodrasi, ``Variational autoencoder for personalized pathological speech enhancement,'' in \emph{Proc. European Signal Processing Conference}, Palermo, Italy, Sept. 2024.

\bibitem{fwSSNR}
Y.~Hu and P.~C. Loizou, ``Evaluation of objective quality measures for speech enhancement,'' \emph{IEEE Transactions on Audio, Speech, and Language Processing}, vol.~16, no.~1, pp. 229--238, 2008.

\bibitem{pesq_2001}
A.~Rix, J.~Beerends, M.~Hollier, and A.~Hekstra, ``Perceptual evaluation of speech quality (pesq)-a new method for speech quality assessment of telephone networks and codecs,'' in \emph{Proc. IEEE Int. Conf. on Acoustics, Speech and Signal Processing (ICASSP)}, vol.~2, 2001, pp. 749--752.

\bibitem{vincent_spectral_2018}
T.~Gerkmann and E.~Vincent, ``Spectral {Masking} and {Filtering},'' in \emph{Audio {Source} {Separation} and {Speech} {Enhancement}}, 1st~ed.\hskip 1em plus 0.5em minus 0.4em\relax Wiley, Sep. 2018, pp. 65--85.

\bibitem{roux_sdr_2019}
J.~L. Roux, S.~Wisdom, H.~Erdogan, and J.~R. Hershey, ``{SDR} – {Half}-baked or {Well} {Done}?'' in \emph{Proc. IEEE Int. Conf. on Acoustics, Speech and Signal Processing (ICASSP)}, May 2019, pp. 626--630.

\bibitem{song2019generative}
Y.~Song and S.~Ermon, ``Generative modeling by estimating gradients of the data distribution,'' in \emph{Proc. NeurIPS}, 2019.

\bibitem{song_2021_score}
Y.~Song \emph{et~al.}, ``Score-based generative modeling through stochastic differential equations,'' in \emph{Proc. Int. Conf. Learning Representations (ICLR)}, May 2021.

\bibitem{richter2023sgmse}
J.~Richter, S.~Welker, J.-M. Lemercier, B.~Lay, and T.~Gerkmann, ``Speech enhancement and dereverberation with diffusion-based generative models,'' \emph{IEEE/ACM Trans. on Audio, Speech, and Language Process.}, vol.~31, pp. 2351--2364, 2023.

\bibitem{chen2023sb}
Z.~Chen, G.~He, K.~Zheng, X.~Tan, and J.~Zhu, ``{Schrodinger Bridges Beat Diffusion Models on Text-to-Speech Synthesis},'' \emph{arXiv preprint arXiv:2312.03491}, Dec. 2023.

\bibitem{os}
F.~Eyben, M.~W{\"o}llmer, and B.~Schuller, ``{openSMILE} -- the munich versatile and fast open-source audio feature extractor,'' in \emph{Proc. ACM International Conference on Multimedia}, Florence, Italy, Oct. 2010, pp. 1459--1462.

\bibitem{eyben2016geneva}
F.~Eyben, K.~R. Scherer, B.~W. Schuller, J.~Sundberg, E.~Andre, C.~Busso, L.~Y. Devillers, J.~Epps, P.~Laukka, S.~S. Narayanan, and K.~P. Truong, ``The geneva minimalistic acoustic parameter set (gemaps) for voice research and affective computing,'' \emph{IEEE Transactions on Affective Computing}, vol.~7, no.~2, pp. 190--202, July 2016.

\bibitem{garofolo_john_s_csr-i_2007_wsj0}
{Garofolo, John S.}, {Graff, David}, {Paul, Doug}, and {Pallett, David}, ``{CSR}-{I} ({WSJ0}) {Complete},'' May 2007.

\bibitem{guevara-rukoz_crowdsourcing_2020}
A.~Guevara-Rukoz, I.~Demirsahin, F.~He, S.-H.~C. Chu, S.~Sarin, K.~Pipatsrisawat, A.~Gutkin, A.~Butryna, and O.~Kjartansson, ``Crowdsourcing {Latin} {American} {Spanish} for {Low}-{Resource} {Text}-to-{Speech},'' in \emph{Proc. {Language} {Resources} and {Evaluation} {Conference}}, Marseille, France, May 2020, pp. 6504--6513.

\bibitem{barker_chime3_2015}
J.~Barker, R.~Marxer, E.~Vincent, and S.~Watanabe, ``The third ‘{CHiME}’ speech separation and recognition challenge: {Dataset}, task and baselines,'' in \emph{{IEEE} {Workshop} on {Automatic} {Speech} {Recognition} and {Understanding} ({ASRU})}, Dec. 2015, pp. 504--511.

\bibitem{lemercier_storm_2023}
J.-M. Lemercier, J.~Richter, S.~Welker, and T.~Gerkmann, ``{StoRM}: {A} {Diffusion}-{Based} {Stochastic} {Regeneration} {Model} for {Speech} {Enhancement} and {Dereverberation},'' \emph{IEEE/ACM Transactions on Audio, Speech, and Language Processing}, vol.~31, pp. 2724--2737, 2023.

\bibitem{torcoli_pesqc2_2025}
M.~Torcoli, M.~M. Halimeh, and E.~A.~P. Habets, ``Navigating {PESQ}: {Up}-to-{Date} {Versions} and {Open} {Implementations},'' May 2025, arXiv:2505.19760 [eess].

\bibitem{cohen_statistical_2009}
J.~Cohen, \emph{Statistical power analysis for the behavioral sciences}, 2nd~ed.\hskip 1em plus 0.5em minus 0.4em\relax New York, NY: Psychology Press, 2009.

\bibitem{itu_p8621}
\BIBentryALTinterwordspacing
ITU-T, ``P.862.1:{Mapping} function for transforming {P}.862 raw result scores to {MOS}-{LQO}.'' [Online]. Available: \url{https://www.itu.int/rec/T-REC-P.862.1-200311-W/en}
\BIBentrySTDinterwordspacing

\bibitem{konane_impact_lang_2021}
D.~Konane, S.~Tiemounou, and W.~Y. S.~B. Ouedraogo, ``Impact of {Languages} and {Accent} on {Perceived} {Speech} {Quality} {Predicted} by {Perceptual} {Evaluation} of {Speech} {Quality} ({PESQ}) and {Perceptual} {Objective} {Listening} {Quality} {Assessment} ({POLQA}): {Case} of {Moore}, {Dioula}, {French} and {English},'' \emph{Open Journal of Applied Sciences}, vol.~11, no.~12, pp. 1324--1332, Dec. 2021, publisher: Scientific Research Publishing.

\end{thebibliography}

\end{document}